\documentclass[useAMS,usenatbib]{mn2e}

\usepackage{natbib}
\usepackage{amsmath,amssymb}
\usepackage[dvips]{graphicx}
\usepackage{xspace}

\usepackage{booktabs}
\usepackage{longtable}
\usepackage{lscape}
\usepackage{array}

\usepackage[english,british]{babel}

\renewcommand{\mathbf}[1]{\boldsymbol{#1}}
\newcommand{\ELAIS}{\textit{ELAIS}\xspace}
\newcommand{\ISOCAM}{\textit{ISOCAM}\xspace}
\newcommand{\ISO}{\textit{ISO}\xspace}
\renewcommand{\micron}{\,$\mu$m\xspace}
\newcommand{\bmicron}{\,$\bmath{\mu}$\textbf{m}\xspace}

\renewcommand{\arcmin}{\hbox{$^\prime$}\xspace}
\renewcommand{\arcsec}{\hbox{$^{\prime\prime}$}\xspace}
\newcommand{\dispra}[3]{$#1^h #2^m #3^s$}
\newcommand{\dispdec}[3]{$#1^\circ #2^\prime #3^{\prime\prime}$}

\newcolumntype{C}{>{\scriptsize}c}
\newcolumntype{L}{>{\scriptsize}l}
\newcolumntype{R}{>{\scriptsize}r}
 
\newcommand\aap{A\&A}
\newcommand\mnras{MNRAS}
\newcommand\aaps{A\&AS}
\newcommand\apj{ApJ}
\newcommand\aj{AJ}
\title[Optical Identifications of \ELAIS sources]{The European Large Area ISO
  Survey (ELAIS): Optical Identifications of 15\bmicron and 1.4\,GHz sources in N1 and N2}

\author[E. A. Gonzalez-Solares et al.]%
{E. A. Gonzalez-Solares$^1$\thanks{E-mail: eglez@ast.cam.ac.uk}, 
I. Perez-Fournon$^2$,
M. Rowan-Robinson$^3$,
S. Oliver$^4$, 
\newauthor
M. Vaccari$^5$,
C. Lari$^6$,
M. Irwin$^1$,
R.G. McMahon$^1$,
S. Hodgkin$^1$,
P. Ciliegi$^7$,
\newauthor
S. Serjeant$^8$,
C.J. Willott$^9$ \\
$^1$Institute of Astronomy, University of Cambridge, Madingley Road, Cambridge
CB3 0HA, UK\\
$^2$Instituto de Astrof\'\i sica de Canarias, C/Via Lactea, s/n, La Laguna
E38200, Spain \\
$^3$Astrophysics Group, Blackett Laboratory, Imperial College of Science,
Technology \& Medicine, Prince Consort Road, London SW7 2BZ \\
$^4$Astronomy Centre, Department of Physics and Astronomy, University of
Sussex, Falmer, Brighton BN1 9QJ, UK \\
$^5$ Dipartamento di Astronomia, Universita\' di Padova, Vicolo Osservatorio 5, I-35122 Padova, Italy \\
$^6$ Istituto di Radioastronomia, Via P. Gobetti 101, Bologna 40129, Italy \\
$^7$ Osservatorio Astronomico di Bologna, Via Ranzani 1, 40127 Bologna, Italy\\
$^8$ Centre for Astrophysics and Planetary Science, School of Physical Sciences, University of Kent, Canterbury, Kent, CT2 7HR, UK\\
$^9$ Herzberg Institute of Astrophysics, National Research Council, 5071 West Saanich Rd, Victoria, B.C. V9E 2E7, Canada
}

\begin{document}

\maketitle

\date{}

\begin{abstract}
  We present the optical identification of mid-IR and radio sources
  detected in the European Large Area ISO Survey (\ELAIS) areas N1 and
  N2. Using the r' band optical data from the Wide Field Survey we
  apply a likelihood ratio method to search for the counterparts of
  the 1056 and 691 sources detected at 15$\mu$m and 1.4\,GHz
  respectively, down to flux limits of $S_{15}=0.5$\,mJy and
  $S_{1.4\,{\rm GHz}}=0.135$\,mJy.  We find that $\sim$92\% of the
  15$\mu$m \ELAIS sources have an optical counterpart down to the
  magnitude limit of the optical data, r'=24.  All mid-IR sources with
  fluxes $S_{15}\geq3$\,mJy have an optical counterpart.  The
  magnitude distribution of the sources shows a well defined peak at
  relatively bright magnitudes r'$\sim$18. About 20\% of the
  identified sources show a point-like morphology; its magnitude
  distribution has a peak at fainter magnitudes than those of
  galaxies.  The mid-IR-to-optical and radio-to-optical flux diagrams
  are presented and discussed in terms of actual galaxy models.
  Objects with mid-IR-to-optical fluxes larger than 1000 are found
  that can only be explained as highly obscured star forming galaxies
  or AGNs. Blank fields being 8\% of the 15$\mu$m sample have even
  larger ratios suggesting that they may be associated with higher
  redshift and higher obscured objects.

\end{abstract}


\section{Introduction}

The Infrared Space Observatory \citep[ISO, ][]{1996A&A...315L..27K}
was the second infrared space mission, providing a great improvement
in sensitivity over the IRAS mission.  The European Large-Area ISO
survey \citep[\ELAIS, ][]{2000MNRAS.316..749O} was the largest Open
Time programme on ISO.  This project surveyed 12 square degrees,
divided into four main fields, three in the north (N1, N2, N3) and
one in the south (S1). The main survey bands used were 6.7, 15, 90 and
170\micron. The ISOCAM camera \citep{1996A&A...315L..32C} was used for
the shorter wavelengths while the ISOPHOT \citep{1996A&A...315L..64L}
camera was used for the longer ones.

%The 6.7$\mu$m survey covers the N2 and N3 regions. The data have
%been processed throughout a Preliminary Analisys and 438 sources
%have been extracted. On the basis of results shown by Vaisanen et al
%(2002) the quality of this analysis is appropriate to derive reliable
%sources and fluxes. 

%The 15\micron survey carried out using
%the \ISOCAM instrument is made from single rasters of
%$\sim43.5^\prime\times42^\prime$ in size. Each of the N1 and N2 fields is
%made from 6 of these rasters, with an integration time of ... per raster. The
%second raster in N1\_2 has been observed twice.

Optical imaging is essential to study the properties of the sources
detected. Due to the large errors ellipses of the mid-IR detections,
typically several seconds of arc, it is necessary to carry out a detailed
process of identification. Since more than one optical source can be
inside those ellipses, a method which provides the likelihood of each
counterpart to be the true association is needed. The identification
does not only provides us the optical properties of the mid-IR sources
but also allows us to further proceed with followup observations of
interesting sources.

This paper presents the optical identification of the mid-IR and radio
sources in the N1 and N2 areas. Section~\ref{sec:wfs} presents a
summary of the optical observations carried out in these areas as well
as the reduction steps and products. Section~\ref{sec:elaiscat}
describes the mid-IR and radio catalogues used.  Section~\ref{sec:id}
discusses the actual procedure to determine the optical counterparts
of the sources, while sections~\ref{sec:properties}
and~\ref{sec:properties2} describe the optical properties of the
sources.

\section{The optical catalogues} \label{sec:wfs}

In order to identify the mid-IR sources with optical objects we use
the data from the Wide Field Survey \citep[WFS,
][]{2001NewAR..45...97M}.  This survey has been carried out using the
Wide Field Camera (WFC) on the 2.5\,m Isaac Newton Telescope (INT) on
the Observatorio del Roque de Los Muchachos (La Palma). The WFC is
formed by 4 4k$\times$2k CCDs. The arrays have 13.5\micron pixels
corresponding to 0.33\arcsec/pixel at the telescope prime focus and
each one covers an area on sky of 22.8$\times$11.4\,arcmin. The total
sky coverage per exposure for the array is therefore 0.29 square
degrees.

\begin{figure}
\centering
\includegraphics[width=0.4\textwidth]{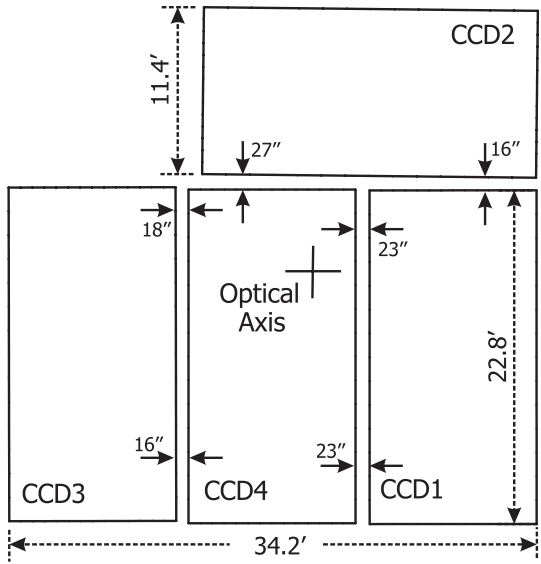}
\caption{Schematic layout of the Wide Field Camera. All the measures are 
approximated.} 
\label{fig:wfc_ccd}
\end{figure}

Figure~\ref{fig:wfc_ccd} shows an schematic layout of the CCD
detectors in the WFC. Gaps between detectors are typically 20\arcsec.
Chip 3 is slightly vignetted in one corner. Optical observations
described below are carried out considering the WFC as a 3-CCD camera,
allowing for a 10\% overlap between adjacent pointings for photometric
purposes. Therefore the area not observed in each pointing due to the
chip gaps is about 12 square arcmin. However, the spatial source
density of \ELAIS and radio sources is not high enough to make this
effect important for the process of identification.

\begin{figure}
  \centering \includegraphics[angle=-90,width=0.98\hsize]{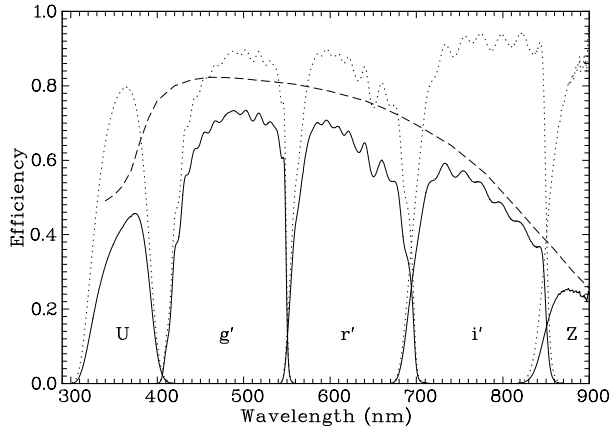}
\caption{Transmission curves of the U, g', r', i' and Z WFC filters 
  (dotted line). Dashed line shows the quantum efficiency of the detector,
  where solid line shows to total throughput of the system.}
\label{fig:filters}
\end{figure}

The WFS surveyed $\sim$200 square degrees in different well known regions of
sky with data at other wavelengths. The \ELAIS regions N1 and N2 were also
included. The survey consists of single 600\,s exposures in five bands: U,
g', r', i' and Z (see figure~\ref{fig:filters}) to magnitude limits of: 23.4,
24.9, 24.0, 23.2, 21.9 respectively (Vega, 5$\sigma$ for a point-like
object), i.e., about 1 magnitude deeper than the Sloan Digital Sky Survey
\citep[SDSS; ][]{2000AJ....120.1579Y}. A total of 108 pointings were done in
N1 and N2, covering a total area of 18 square degrees. Typical seeing is
about 1.0-1.2\arcsec. The data are processed by the Cambridge Astronomical
Survey Unit (CASU) as described in Irwin \& Lewis (2000) and we provide here
a short description of the reduction steps. The data are first debiassed
(full 2D bias removal is necessary). Bad pixels and columns are then flagged
and recorded in confidence maps, which are used during catalogue generation.
The CCDs are found to have significant non linearities so a correction using
look-up-tables is then applied to all data.  Flatfield images in each band
are constructed by combining several sky flats obtained in bright sky
conditions during the twilight. Exposures obtained in the i' and Z bands show
a significant level of fringing ($\pm$2\% and $\pm$6\% of sky respectively).
In order to remove this effect, master fringe frames are created by combining
all the science exposures for each band.  These fringe frames are then
subtracted from the object exposures. After this removal, the fringing level
is reduced to $\pm$0.2\% and $\pm$0.4\% of sky in the i' and Z bands
respectively.  Finally an astrometric solution starts with a rough WCS based
on the known telescope and camera geometry and is the progressively refined
using the Guide Star Catalogue for a first pass and the APM or PMM catalogues
for a final pass. The WFC field distortion is modelled using a zenithal
equidistant projection \citep[ZPN; ][]{2002AA...395.1061G}. The resulting
internal astrometric precision is better than 100 mas over the whole WFC
array (based on intercomparison of overlap regions).  Global systematics are
limited by the precision of the APM and PMM astrometric catalogue systems and
are at the level of 300 mas. The object detection is performed in each band
separately using a standard APM-style object detection and parametrisation
algorithm. Standard aperture fluxes are measured in a set of apertures of
radius $r/2$, $r$, $\sqrt{2}\,r$, $2\,r$, $2\,\sqrt{2}\,r$ where $r=3.5$
pixels and an automatic aperture correction (based on the average
curve-of-growth for stellar images) is applied to all detected objects.

Photometric calibration is done using series of Landoldt standard
stars (Landoldt 1992) with photometry in the SDSS system. For each
night a zero point in each filter is derived. For photometric nights
the calibration over the whole mosaic has an accuracy of 1-2\%. During
non-photometric nights, in otherwise acceptable observing conditions,
we find that the derived zeropoint systematic errors can be up to 10\%
or more.  Although the pipeline usually successfully flags such nights
as non-photometric it still leaves open the problem of what to do
about tracking the varying extinction during these nights.

All calibration is by default corrected for the mean atmospheric
extinction at La Palma during pipeline processing (0.46 in U, 0.19 in
g$\arcmin$, 0.09 in r$\arcmin$ and 0.05 in i$\arcmin$ and Z).  Since
adjacent camera pointings overlap by several square arcminutes sources
in these overlapping regions can be used as magnitude comparison
points.  However, to use overlapping and hence in general, different
CCDs, requires that any repeatable systematics due to, for example,
slight differences in the colour equations for each CCD, are first
corrected for.  Correcting for these is a three stage process.  First
the twilight flatfields are used to gain-correct each CCD onto a
common system.  However, since the twilight sky is significantly bluer
than most astronomical objects, a secondary correction is made using
the measured dark sky levels in each CCD for each filter to provide a
correction more appropriate for the majority astronomical object.
These corrections, unsurprisingly, are negligible for passbands on the
flat part of the generic CCD response curves such as g' and r', and
amount to 1-2\% for the i' and z' passbands.  The measured dark sky
values for the U-band were also consistent with zero correction though
with less accuracy due to the low sky levels in the U-band images.
Finally any residual offsets between the CCDs are checked for each
survey filter using the mean offset between adjacent pointings on
photometric survey nights.  The only filters requiring significant
adjustments to the individual CCD zero points at this stage are CCD3
for U (-3\%) and CCD1 for z' (+3\%).

Data from non-photometric nights can now be calibrated in one of two
ways: the overlap regions between pointings can be used to directly
tie in all the frames onto a common system, with extra weighting given
to data taken from photometric nights; the stellar locus in various
2-colour diagrams (in regions of low unchanging extinction like these)
can be used to compare colours from all the passbands by
cross-correlating the loci between any of the pointings.  The
colour-colour loci cross-correlations (actually made from a smoothed
Hess-like version of the diagrams) give results accurate to better
than +/-3\% and can be used in conjunction with the overlap results to
give overall photometry for the survey to the level of $\approx$2\%.
The final products include astrometrically calibrated images as well
as morphologically classified merged multicolour catalogues,
publically available from the WFS web page
(\verb+http://www.ast.cam.ac.uk/~wfcsur/index.php+).

%A zero point averaged over each run is also derived and used in cases
%when insufficient standards were observed in a particular night. This
%method was found to be accurate to a $\sim$10\% level in the \ELAIS
%regions. In order to improve this calibration, advantage of the
%overlapping regions was taken into account in a further step. [Provide
%some text about how this was done.]  The resulting photometry is
%accurate in all bands within 2\% error.

For the purpose of this work, we have also carried out the object
detection in the r' band images using
SExtractor~\citep{1996AAS..117..393B}. As well as providing a
independent check for the extraction method used above, it gives us an
extense set of parameters. Aperture magnitudes from SExtractor are
found to be in agreement with the ones obtained from the WFS pipeline.
Therefore, together with the WFS aperture magnitudes we use
\texttt{MAG\_BEST} as a measurement of the total magnitude in the
r'-band. The galaxy-star classification given by \texttt{CLASS\_STAR}
is also used to select point-like objects (defined to be
\texttt{CLASS\_STAR} $>$ 0.8).

\begin{figure}
\includegraphics[width=0.45\textwidth,clip]{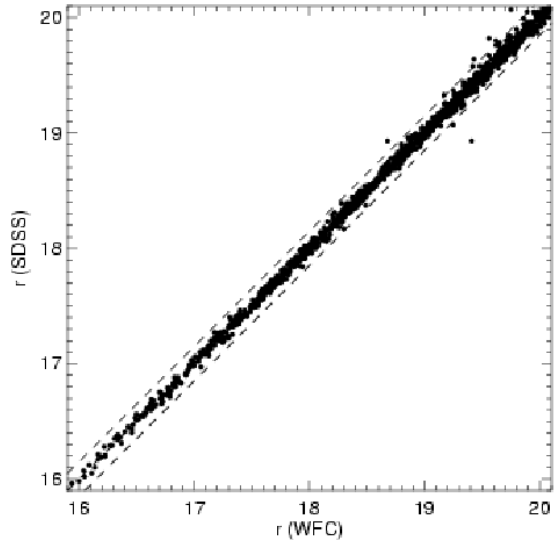}
\caption{Comparison between WFS and SDSS r' band magnitudes for point-like 
  objects. A correction $\Delta m = 0.09$ has been applied to account
  for the slight filter and system differences.}
\label{fig:wfs_sdss_r}
\end{figure}

As an additional test of our photometric calibration we have
correlated the WFS catalogues with those from the Sloan Digital Sky
Survey First Data Release \citep{2003astro.ph..5492R}.
Figure~\ref{fig:wfs_sdss_r} shows the result of this correlation for
the r' band magnitude. Once accounted for the small correction between
both filters ($\Delta m = 0.09$) the agreement is within 0.04
magnitudes (note also that a second order correction due to different
object spectral energy distributions is not removed, so the accuracy
is better than the quoted 0.04).

\section{The mid-IR and radio catalogues} \label{sec:elaiscat}

The final analysis of the 15$\mu$m data using the Lari method
\citep{2001MNRAS.325.1173L} for the \ELAIS northern fields has recently
been completed (Maccari et~al. 2004, in prep). The two main northern
fields N1 and N2 are centred at \dispra{16}{10}{01}
\dispdec{+54}{30}{36} and \dispra{16}{36}{58} \dispdec{+41}{15}{43}
% and \dispra{14}{29}{06} \dispdec{+33}{06}{00}
respectively. We have obtained a sample of 1056 sources (490 in N1 and 566 in
N2) to a $5\sigma$ flux limit of 0.45\,mJy. This catalogue includes sources
detected in deeper observations (in particular, the central
40\arcmin$\times$40\arcmin N2 area has been observed three times) so this
flux limit is not homogenous over the whole survey area.
%As shown by Maccari et al,, the completeness of the
%catalogue is 90\% at 1\,mJy and reduces to 50\% at ??\,mJy.

As part of the multiwavelenth follow-up observations carried out in
these regions \cite{1999MNRAS.302..222C} have conducted a survey at
20\,cm using the VLA in its C configuration, covering 4.22 sq. deg. in
the N1, N2 and N3 areas. They detect a total of 867 sources (362 in
N1, 329 in N2 and 176 in N3) above a flux limit 0.135 mJy in the
deeper observations or 1.15 mJy over the shallower ones. We have
selected sources for which we have optical data from the WFS (i.e.,
sources in N1 and N2). Multi-component sources (flagged in the
catalogue as 'A', 'B' or 'C' components) have been removed and only
the calculated central position considered ('T' in the catalogue).

\section{Optical identifications of \ELAIS sources} \label{sec:id}

As shown in \cite{2001MNRAS.325.1173L}, the positional errors
in RA and DEC for the \ELAIS sources result from the combination of
three quantities: the finite spatial sampling ($\sigma_{\rm s}$), the
reduction method ($\sigma_{\rm r}$) and the pointing accuracy
($\sigma_{\rm P}$). The simulation work carried out in the \ELAIS fields
yield next relations:
\begin{gather}
\sigma_{\rm s+r}({\rm RA})= 1.00+17.17\times{\rm e}^{(-0.57\times {\rm S/N})}\\
\sigma_{\rm s+r}({\rm DEC})= 1.06+1.21\times{\rm e}^{(-0.16\times {\rm S/N})}
\end{gather}

These equations have been used to estimate the positional errors due to
mapping and reduction method as a function of each source's signal-to-noise
ratio, S/N.

The errors introduced by uncertainties in the \ISOCAM pointing have
been estimated by correlating the \ISO sources with the USNO catalogue
of optical objects \citep{1998AAS...19312003M}.  For each raster, both
catalogues have been correlated using a maximum search distance of
12\arcsec. The \textit{median} of the offsets values for all the
ISO-USNO associations have been calculated,$\sigma_{\rm P}({\rm RA})$,
$\sigma_{\rm P}({\rm DEC})$.  Each source position has then been
corrected for the offset found for each raster. The final positional
error is then:
\begin{gather}
\sigma_{\rm RA}^2=\sigma_{\rm s+r}^2({\rm RA})+\sigma_{\rm P}^2({\rm RA}) +
0.4^2\\
\sigma_{\rm DEC}^2=\sigma_{\rm s+r}^2({\rm DEC})+\sigma_{\rm P}^2({\rm DEC})
+ 0.4^2
\label{eq:sigmas}
\end{gather}
where a $\sigma=0.4$\arcsec has been added to account for the optical
errors. Typical positional errors are $\sigma\sim3$\arcsec.

\begin{figure}
\includegraphics[width=0.235\textwidth]{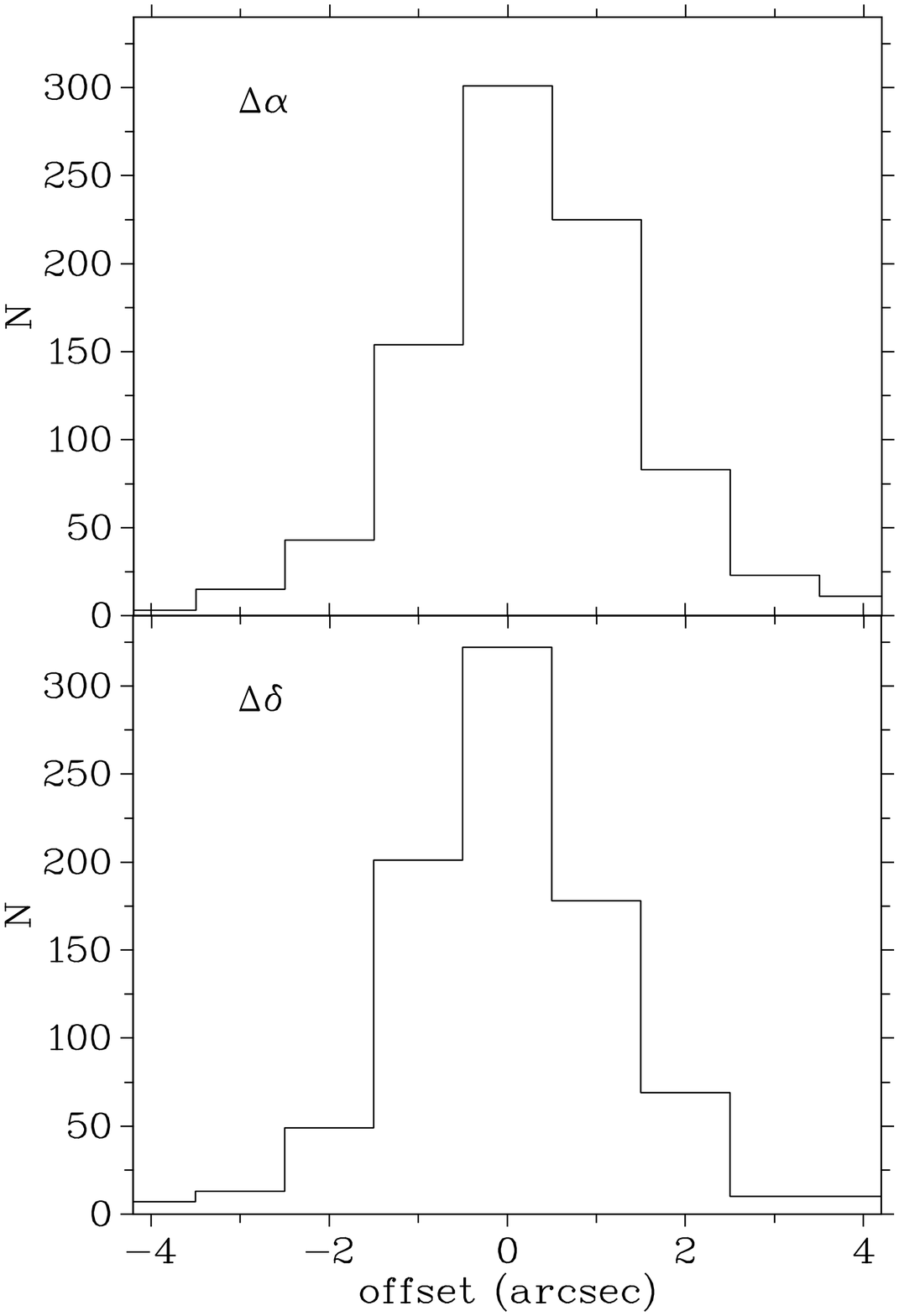}
\includegraphics[width=0.235\textwidth]{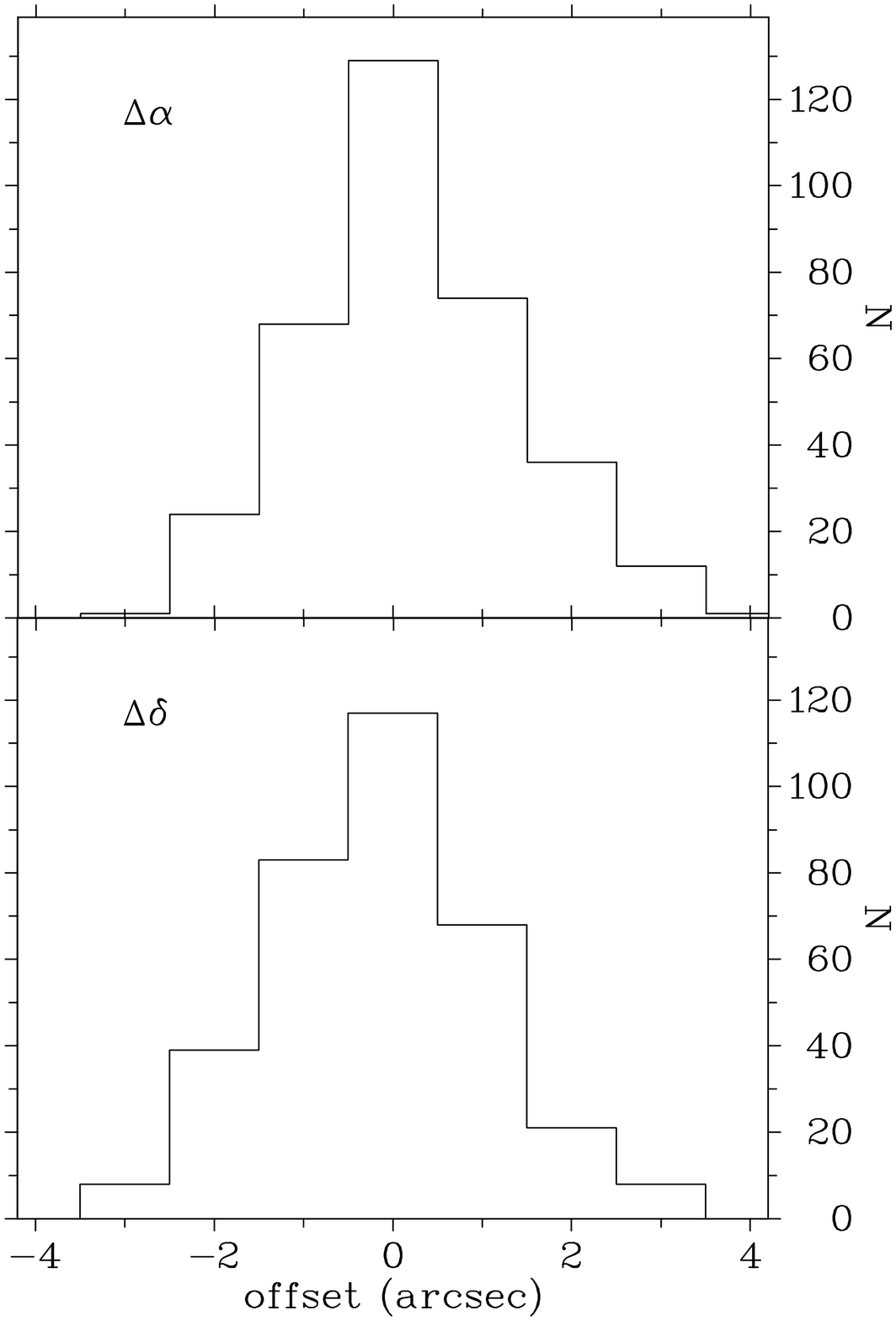}
\caption{Offsets between 15$\mu$m and optical positions (left) and
between radio and optical positions (right).}
\label{fig:dist}
\end{figure}

The correlation between \ELAIS 15$\mu$m sources and optical objects has been
carried out using a likelihood ratio method \citep{1992MNRAS.259..413S},
similar to the one which has been successfully applied to the identification
of 15$\mu$m sources detected by ISO in the HDF-N by
\cite{1997MNRAS.289..482M}. 

The probability that an optical object of magnitude $m$ is the true
counterpart of a source with an error ellipse defined by its major axis,
$\sigma_1$, and minor axis, $\sigma_2$, separated a distance $r$ is given by
\begin{equation}
\mathcal{L} = \frac{Q(m) exp (-r^2/2)}{2\pi\sigma_1\sigma_2 N(m)}
\label{eq:like}
\end{equation}
where $Q$ and $N$ are the magnitude distributions of the sources
and objects respectively. The reliability of such identification is
\begin{equation}
\mathcal{R}_j = \frac{ \mathcal{L}_j } {\sum_i \mathcal{L}_i + (1 - Q)}
\label{eq:rel}
\end{equation}
The identification process is carried out as follows. For each \ELAIS 15$\mu$m
source, all optical objects within a distance of 20\arcsec are selected. This
list is our candidate list. For each object in our candidate list we
calculate the values of the likelihood and reliability as given by equations
above. The likelihood value (equation \ref{eq:like}) gives us the probability
that a candidate is the true optical counterpart of the source; but it only
provides information about the probability of each candidate being the
correct counterpart. The reliability (equation \ref{eq:rel}) provides
information about the number of candidates with high likelihood values. A
candidate will have large values of likelihood and reliability if it is the
only probable counterpart of a source. In case where there are multiple
probable counterparts (in the sense of high likelihood), they all will have
low values of reliability.

A candidate is selected to be the correct optical identification of a
\ELAIS source when $\mathcal{L} > 0.8$. Sources for which no candidates
meet this requirement are flagged as blank fields and represent
$\sim8\%$ of the total sample (section~\ref{sec:blanks}). Sources for
which there are more than one candidate meeting this requirement, and
have low values of reliability, are flagged as having multiple
possible counterpart and represent $\sim8\%$ of the sample.  Finally
bright stars, saturated in the WFS CCD data are also flagged; note
that their astrometric accuracy is poor. They represent $\sim22\%$ of
the sample.  Figure~\ref{fig:dist} (left) shows the offsets between
ISO and optical identifications, excluding saturated stars.
Uncertainties are well fitted by a Gaussian distribution of
$\sigma\sim1$\arcsec.

Optical identification of the \ELAIS radio sources detected in the N1
and N2 areas is carried out using a similar procedure. In this case
the radio catalogue provides measurements of the positional errors, so
these are used in our likelihood ratio algorithm.  An optical
counterpart is found for 389 out of the 691 sources, i.e., there is a
44\% of blank fields. Figure~\ref{fig:dist} (right) shows the offsets
between radio sources and the optical counterpart. This provides a
confirmation of the good accuracy of the astrometry of the 15\micron
sources.

\begin{figure*}
\includegraphics[width=0.95\textwidth]{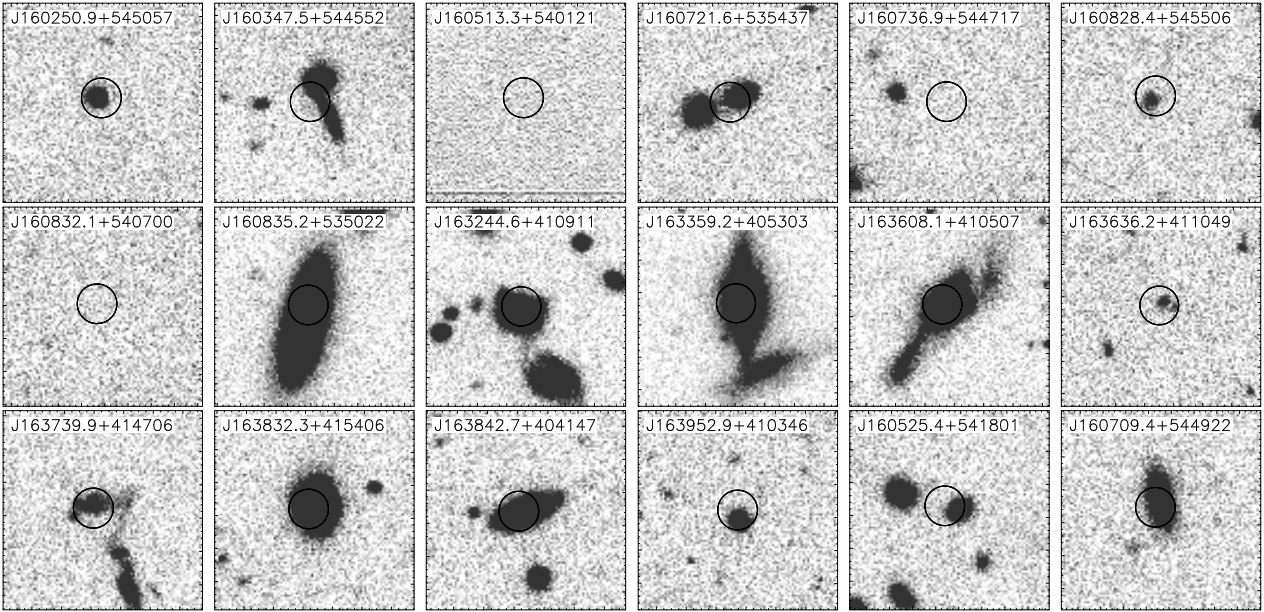}
\caption{Example r'-band postage stamps of 15$\mu$m detected sources. 
  Each chart is 30\arcsec$\times$30\arcsec in size.}
\label{fig:fcharts}
\end{figure*}

\begin{figure}
\centering \includegraphics[width=0.9\hsize]{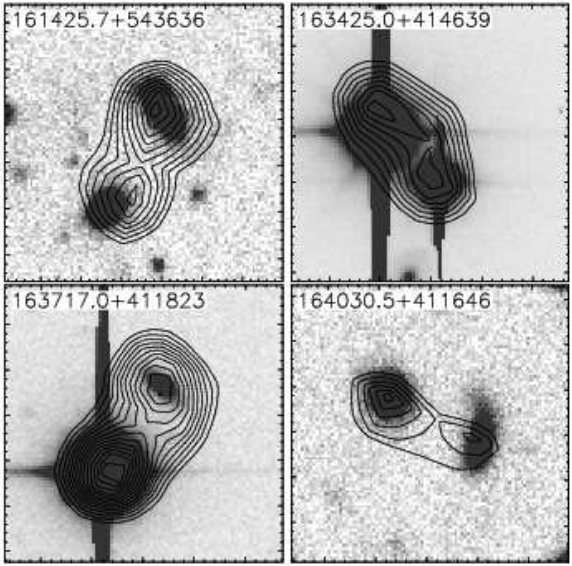}
\caption{Double sources merged in the \ELAIS 15$\mu$m catalogue. 
  Mid-IR contours starting at $5\sigma$ and increasing at steps of
  $1\sigma$ are shown plotted over the r'-band image.  Each chart is
  30\arcsec$\times$30\arcsec in size.}
\label{fig:doubles}
\end{figure}

Figure~\ref{fig:fcharts} shows some example finding charts of sources
detected at 15$\mu$m. These charts are 30\arcsec$\times$30\arcsec in
size and have been extracted from the r'-band images. Also shown a
3\arcsec radius circle centered on the ISO position. As shown in
section~\ref{sec:properties} a large fraction of the sources are
associated with bright galaxies.  Example of sources with more than
one plausible counterpart is given by e.g., J163636.2+411049. The most
likely counterpart is an object with r'=21.8 at a distance of 1.1
arcsec from the ISO position. Its likelihood is 0.988 and reliability
is 0.773.  There is another source of r'=23.0 separated 2 arcsec from
the ISO position with likelihood 0.960 and reliability 0.227. We
choose the first one as the optical counterpart (based on higher
values of likelihood and reliability) of the ISO source and flag it
has having more than one plausible counterpart. Examples of these are
also J160347.5+544552 and J160721.6+535437. 

There are four \ELAIS sources which have been merged in the final
analysis 15\,$\mu$m catalogue (figure 6). Both possible counterparts
are given in the optical identification catalogue and the 15\,$\mu$m
flux is assigned to both.

The table of optical identifications is available electronically in:
\verb+http://www.ast.cam.ac.uk/~eglez/eid+. Also included are
postage stamps for all the sources in the r' band (grayscale and
contours) as well as multiband grayscale finding charts.

The format of the table is as follows:

\begin{description}
\item[Column 1.] International Astronomical Unit name of the
source. Sources are listed in RA order. Sources detected in N2 are
listed after those detected in N1.
\item[Column 2.] ISO coordinates (J2000) of the sources. 
\item[Column 3.] Coordinates (J2000) of the optical counterpart of the
  source.
\item[Columns 4 to 8.] Aperture magnitude in U, g', r', i' and Z bands
  (aperture of radius 3.5 pixels -- 1.16\arcsec).
\item[Column 9.] Total r'-band magnitude as provided by SExtractor
  \texttt{MAG\_BEST} parameter.
\item[Columns 10 to 15.] Errors in previous magnitudes.
\item[Columns 16 to 21.] Stellar classification as provided by WFS and
  SExtractor \texttt{CLASS\_STAR} parameters.
\item[Columns 22, 23 and 24.] Distance, $\Delta\alpha$ and
$\Delta\delta$ between the \ELAIS source and the optical association.
\item[Columns 25 and 26.] Likelihood of the identification formated as
  $\mathcal{L}/(1+\mathcal{L})$ and reliability.
\item[Columns 27 and 28.] Flux at 15$\mu$m and signal-to-noise ratio.
\item[Column 29.] Optical flag code as follows. B1: source in a gap
  between chips or in the edge of a chip, B3: most likely optical
  identification when multiple counterparts, B4: blank field, B7: bright
  saturated star, B9: other plausible optical identification when
  multiple counterparts.
\end{description}

\section{Magnitude distributions}
\label{sec:properties}

\begin{figure}
\centering
\includegraphics[angle=-90,width=0.45\textwidth, clip=true]{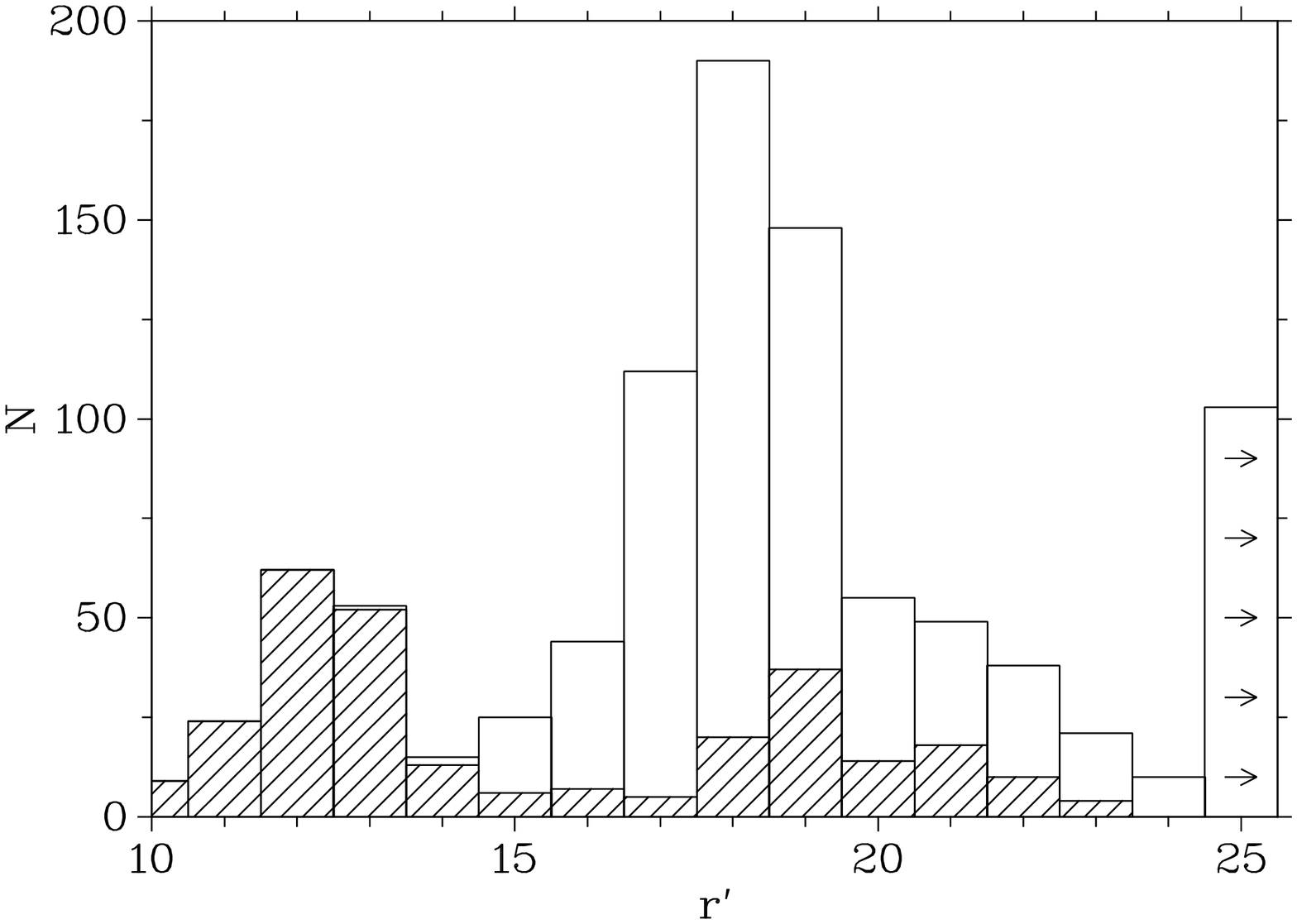} %\quad
\caption{Magnitude distribution of 15\micron sources in \ELAIS N1 and
N2. Hatched histogram shows the distribution of point-like
sources. Last bin, with right arrow symbols, represent the objects
without optical counterpart.}
\label{fig:magdist_iso}
\end{figure}

Figure~\ref{fig:magdist_iso} shows the magnitude distribution of the optical
counterparts of \ELAIS 15\micron sources. The first peak at r'$\sim$12-13 is
caused by bright stars which are saturated in the WFS data. The second peak,
due to extra-galactic objects, is located at r'$=18$. Most of the
extragalactic sources are then associated with optically bright objects. A
tail of faint objects is present at r'$>$20.

In order to calculate the percentage of chance associations, especially at
the faintest magnitudes, we have simulated four catalogues in each N1 and N2
regions by offsetting 20 arcsec in R.A. and DEC from the ISO position in four
directions. The likelihood ratio technique was used to associate the sources
in these new catalogues in the same way as done for the real ones. The
identifications obtained provides the distribution of chance associations.
This percentage is $\sim$5\% for r'$\leq$20 and increases to 20\% at r'=24.

%Shaded histogram
%in figure~\ref{fig:magdist_iso} (top) shows the magnitude distribution of this
%``chance'' identifications. The percentage of false associations is $\sim$5\%
%for r$\leq$20 and increases to 13\% at r=21, 30\% at r=22, ad 77\% at r=23.

Figure~\ref{fig:magdist_iso} also shows the magnitude distribution
of point like objects. About 16\% of the objects with
magnitude r'$>$15 are classified as point like. Their magnitude
distribution has a peak at r'$\sim$19, a magnitude fainter than that
for galaxies.

\begin{figure}
\centering
\includegraphics[angle=-90,width=0.45\textwidth, clip=true]{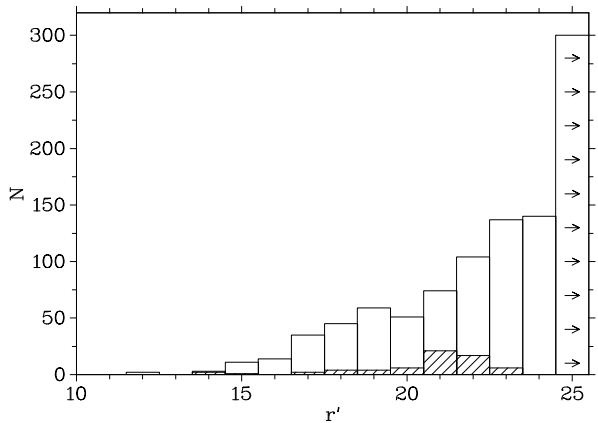} %\quad
\caption{Magnitude distribution of 1.4\,GHz radio sources in \ELAIS N1
  and N2. Hatched histogram shows the distribution of point-like
  sources. Last bin, with right arrow symbols, represent the objects
  without optical counterpart.}
\label{fig:magdist_vla}
\end{figure}

The magnitude distribution of radio sources is shown in
figure~\ref{fig:magdist_vla}. Unlike the distribution of mid-IR
sources the number of sources show an increase at fainter
magnitudes. The number of point-like objects is very low (hatched
histogram) but the reliability of the \texttt{CLASS\_STAR} parameter in
SExtractor decreases at faint magnitudes and fails at r'$\geq$23.

\section{Optical to infrared fluxes}
\label{sec:properties2}

%% Spectroscopic follow-up observations of a sample of $15\mu$m objects
%% has already been carried out (Perez-Fournon et al 2003). Redshifts
%% have been obtained for 130 objects which represent about 15\% of the
%% sample of extragalactic objects. Most (~??\%) of the sample show line
%% ratios typical of star-forming galaxies with a median redshift of 0.2
%% (see figure...).

\begin{figure}
\centering
\includegraphics[angle=-90,width=0.48\textwidth, clip=true]{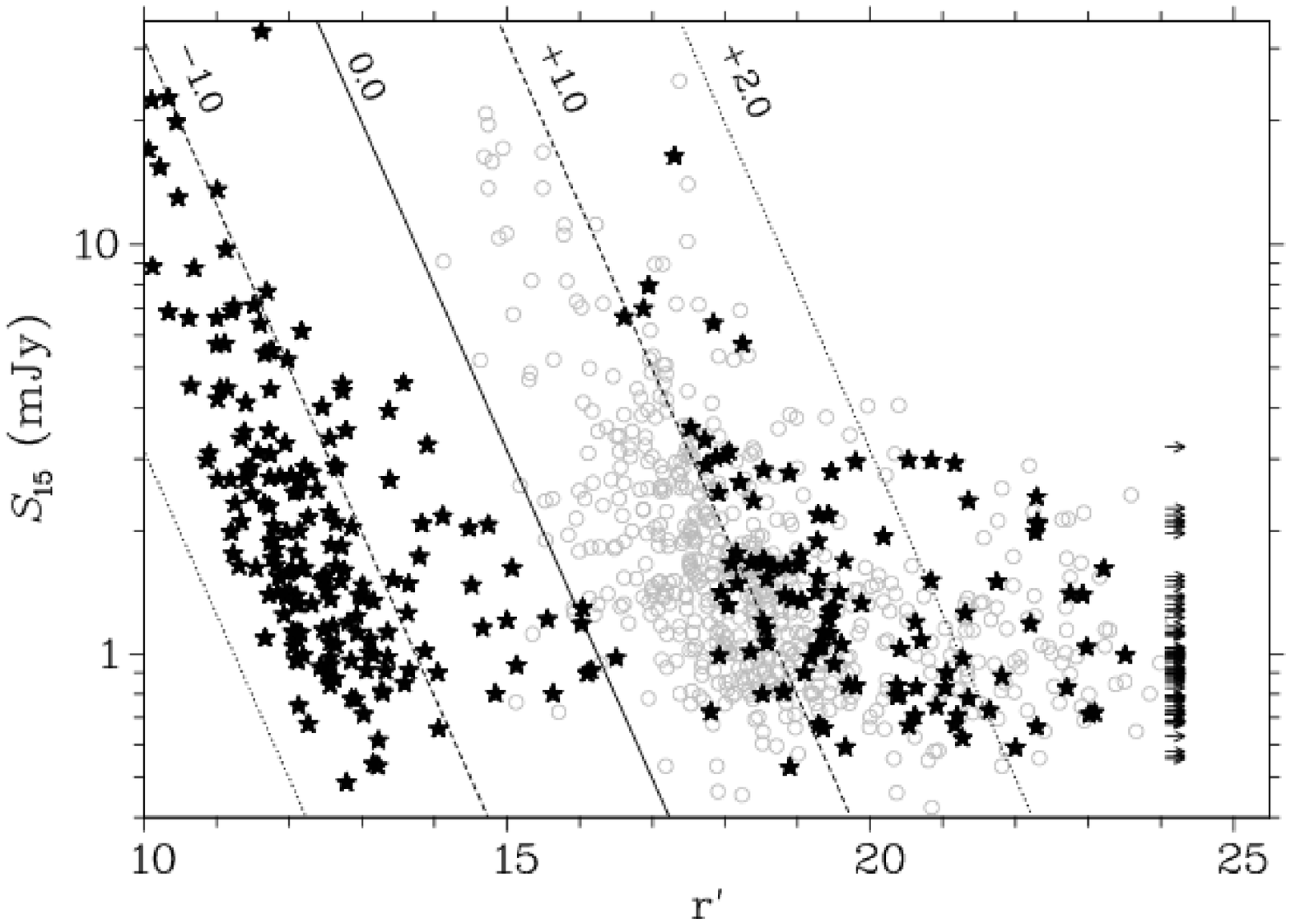} \quad
\includegraphics[angle=-90,width=0.48\textwidth,clip=true]{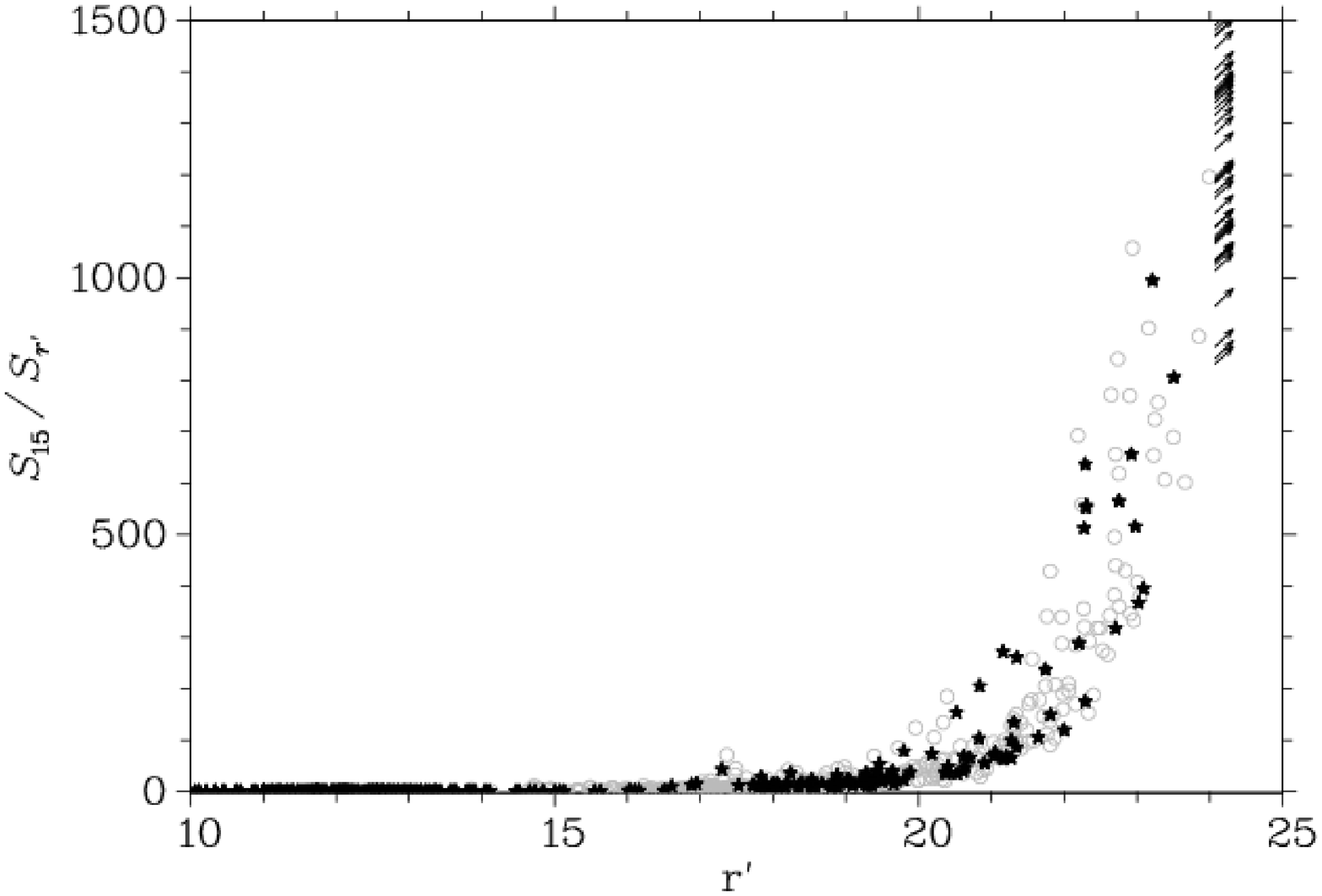}
\caption{Optical to infrared flux ratio for 15$\mu$m sources with
optical counterpart. Top figure shows the mid-IR flux versus optical
magnitude. Point-like classified object are shown as black stars while
galaxies are shown as gray circles. Lines of constant flux ratio
$\log(S_{15}/S_{r'})=-1.0, 0.0, +1.0, +2.0$ are also shown. Blank
fields are displayed using a right arrow. Bottom figure shows
$\log(S_{15}/S_{r'})$ versus optical magnitude. }
\label{fig:s15_rmag}
\end{figure}

Figure~\ref{fig:s15_rmag} shows the mid-IR to optical fluxes for the \ELAIS
sources. Stars have typically low mid-IR fluxes compared to their optical
fluxes and are located in the region where their mid-IR flux is ten times
smaller than their optical flux. Most of the extragalactic objects have
mid-IR fluxes between 1 and 100 times their optical fluxes. According to
models of infrared galaxies previously published \cite{2001NewAR..45..631R}, 
galaxies whose infrared emission is dominated by cirrus are located in
the region $0<\log S_{15}/S_{r'}<1$. More infrared ``active'' galaxies, i.e.,
starbursts, AGN and Arp220-like objects are located in regions of mid-IR flux
10 times to 100 times larger than their optical flux.

\begin{figure}
\centering
\includegraphics[angle=-90,width=0.45\textwidth, clip=true]{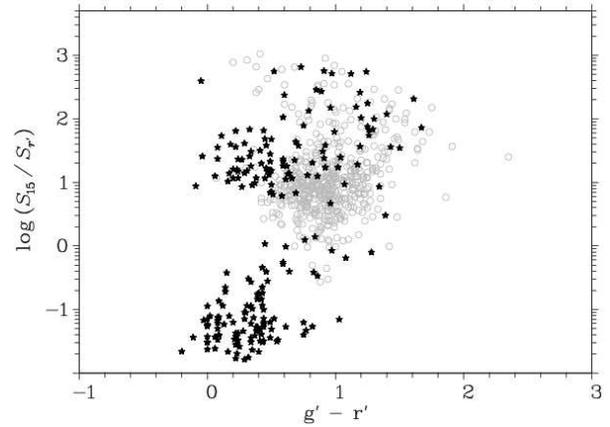}
\caption{Infrared to optical flux ratio versus optical colour $g'-r'$. Symbols as in figure \ref{fig:s15_rmag}.}
\label{fig:s15r_gr}
\end{figure}

Optical colours can be used to discriminate between AGNs and
galaxies. As figure~\ref{fig:s15r_gr} shows AGNs are typically 0.5
magnitudes bluer than galaxies (and as shown in ~\ref{fig:s15r_gr},
about 1 magnitude fainter). There is a population of point-like objects
with galaxy-like colours, presumable highly obscured. Their overall
spectra can be well fitted by a galaxy SED \citep{2003astro.ph..8283R}.

\begin{figure}
\centering
\includegraphics[angle=-90,width=0.45\textwidth, clip=true]{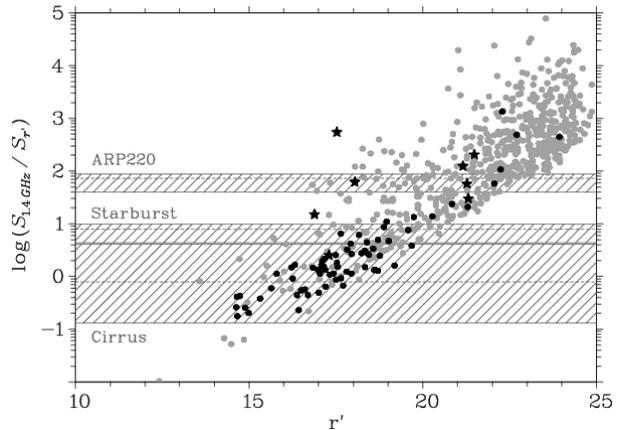}
\caption{Radio to optical flux ratio versus optical magnitude. Black
  symbols represent the sources detected also at 15$\mu$m (dots:
  galaxies, stars: point-like objects). Hatched areas represent the range
  predicted by different infrared galaxy types in the range $0<z<1.5$
  (redshift increases from bottom to top in each area; dashed line shows the
  location of $z=1.0$).}
\label{fig:s20r_rmag}
\end{figure}

\begin{figure}
\centering
\includegraphics[angle=-90,width=0.45\textwidth, clip=true]{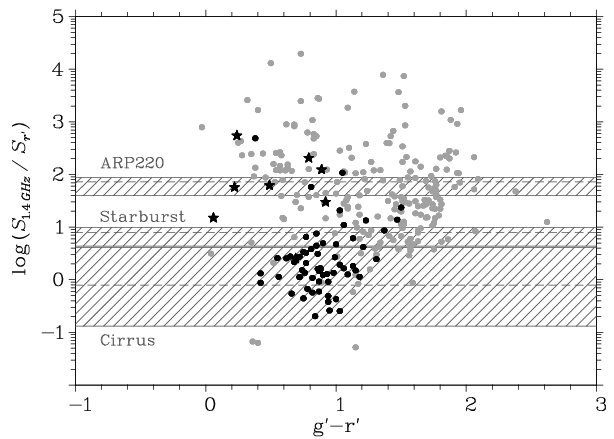}
\caption{Radio to optical flux ratio versus optical colour. Symbols as in
  figure \ref{fig:s20r_rmag}.}
\label{fig:s20r_gr}
\end{figure}

The radio to optical flux ratio versus magnitude and versus optical
colour are shown in figures \ref{fig:s20r_rmag} and \ref{fig:s20r_gr}
for the 1.4\,GHz sources with optical counterpart. Those sources which
show emission at 15$\mu$m are also marked with black symbols.  Most of
the 15\micron - radio coincidences are objects with magnitudes
15$<$r'$<$20 for which the cirrus component is the most plausible source
of the infrared and radio emission.

\begin{figure}
\centering
\includegraphics[angle=-90,width=0.45\textwidth, clip=true]{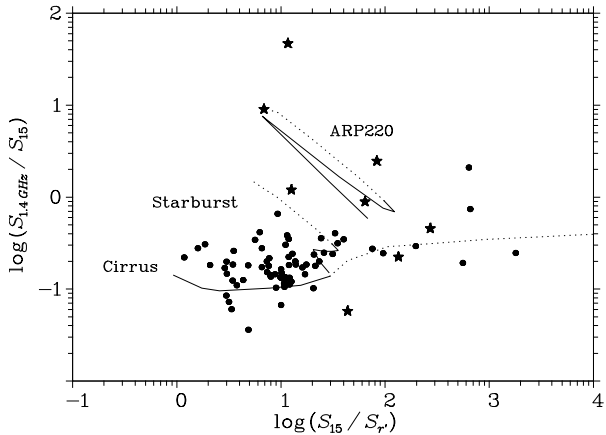}
\caption{Radio to mid-IR flux ratio versus mid-IR to optical flux
  ratio. Symbols as in figure \ref{fig:s20r_rmag}. Lines show the location of
  different infrared galaxy models..}
\label{fig:s20_15_rmag}
\end{figure}

Attending to their radio, mid-IR and optical properties (see
figure~\ref{fig:s20_15_rmag}) the bulk of the 15\micron sources (those
with $0<\log S_{15}/S_{r'}<2$) can be explained as a mixture of
cirrus-dominated galaxies and starburst with smaller fractions of
Arp220-like objects and AGN. 

\subsection{Population with large infrared-to-optical flux ratio}

Figure ~\ref{fig:s15_rmag} also shows the existence of a population of
objects with extreme mid-IR to optical fluxes and faint optical magnitudes.
The nature of this population can only be studied with detailed spectroscopic
followup observations. Two of the point like objects in this region of the
diagram have spectroscopic redshifts.  The object associated with
ELAISC15\_J164021.5+413925 has a 15$\mu$m flux of $S_{15}=2.98$ mJy, an
optical magnitude of r'$=20.3$ and a redshift $z=0.60$ (Perez-Fournon et~al.
2003, in prep.). Its luminosity is $L=10^{11}\,L_\odot$.  The second object is
associated with ELAISC15\_J163655.8+405909. It has a 15$\mu$m flux of
$S_{15}=0.72$ mJy, an optical magnitude of r'$=23.9$ and a redshift $z=2.61$
\citep{2003MNRAS.339..397W}. Its luminosity is $L=10^{12}\,L_\odot$ ($H_0=65,
\Omega_m=0.3, \Lambda_0=0.7$).
%As shown
%in \citeauthor{2003MNRAS.339..397W}, this objects is also very
%reddened.

%In order to explain this population we have used infrared galaxy
%models from \cite{2001ApJ...562..179X}. Two models have been selected,
%corresponding to starburst galaxy and AGN, both with large infrared
%luminosities. Reddening due to star formation has been introduced as
%modeled by \cite{2000ApJ...533..682C}.

Using infrared models of starburst galaxies and AGN, and adding
reddening (as modeled by \cite{2000ApJ...533..682C}), we can explain
the nature of this population. Objects with $\log S_{15}/S_{r'}>2$ can be
explained as luminous starburst galaxies, with luminosities
$L\sim10^{12}\,L_\odot$, at redshifts $z\sim0.7$ and reddening
$A_v\sim1-1.5$. Fainter optical objects with $\log S_{15}/S_{r'}\sim3$
have typically larger luminosities $L>10^{13}\,L_\odot$, redshifts
$z\sim1.3$ and reddening $A_v\sim2$.  Luminous AGN, with
$L\sim10^{12}\,L_\odot$, also populate this area.

The association of this sources with very reddened objects supports
the assumption that the objects with $S_{15}/S_{r'} > 100$ in
figure~\ref{fig:s15_rmag} may represent a new population of heavily
obscured starbursts and type 2 AGN.

%\begin{figure}
%\centering
%\includegraphics[angle=-90,width=0.45\textwidth, clip=true]{S20_15_rmag}
%\caption{Radio to mid-IR flux ratio versus optical magnitude.}
%\label{fig:s20_rmag}
%\end{figure}

\section{Blank fields} \label{sec:blanks}

A percentage of 8\% of the 15$\mu$m sources (38 in N1, 67 in N2) do not have
an optical counterpart down to the optical limits of data described in
previous sections. All of them have infrared fluxes $S_{15} < 3$\,mJy, and
only 3 show a plausible association with a radio source
(table~\ref{tab:blank_radio}). Blank fields are probably the extreme version
of the objects found at $\log S_{15}/S_{r'}>2$. Using the same model galaxies
as before, we find that starburst galaxies with luminosities
$L\sim10^{13}\,L_\odot$, $z\sim1.$ and $A_v\sim3-4$ may populate this region.
The cause of the infrared emission and the starburst activity may be the
merger of galaxies ir the formation of a protogalaxy although the accuracy of
this hypotheses cannot be tested with these data.

%% Recently the \ELAIS N1 regions has been observed by 

\begin{table}
\begin{center}
\caption{\ELAIS blank fields with a radio source nearest than 3\arcsec from the ISO position. Fluxes are given in mJy.} \label{tab:blank_radio}
\begin{tabular}{@{\,\,}L@{\,\,\,\,}C@{\,\,\,\,}C@{\,\,\,\,}C@{\,\,}} \hline\hline
\textbf{Name} & \textbf{ISO Coords (J2000)} & \textbf{$S_{15\mu\rm m}$} & \textbf{$S_{20 \rm cm}$}  \\ \hline
J160734.3+544216 & 16 07 34.40 +54 42 15.6  & 1.51$\pm$0.19 & 0.37$\pm$0.02 \\
J163505.4+412508 & 16 35 05.71 +41 25 11.2  & 0.83$\pm$0.09 & 0.38$\pm$0.02 \\
J163511.4+412255 & 16 35 11.54 +41 22 57.4  & 0.72$\pm$0.09 & 0.94$\pm$0.02 \\ \hline
\end{tabular}
\end{center}
\end{table}

%% \begin{table*}
%% \begin{center}
%% \caption{\ELAIS blank fields with a radio source nearest than 0.3\arcsec from the ISO position. Fluxes are given in mJy.} \label{tab:blank_radio}
%% \begin{tabular}{lcccc} \hline\hline
%% \textbf{Name} & \textbf{ISO Coords (J2000)} &\textbf{VLA Coords (J2000)} & \textbf{$S_{15\mu\rm m}$} & \textbf{$S_{20 \rm cm}$}  \\ \hline
%% J160734.3+544216 & 16 07 34.30 +54 42 16.9 & 16 07 34.40 +54 42 15.6 & 1.51$\pm$0.19 & 0.37$\pm$0.02 \\
%% J163505.4+412508 & 16 35 05.47 +41 25 08.9 & 16 35 05.71 +41 25 11.2 & 0.83$\pm$0.09 & 0.38$\pm$0.02 \\
%% J163511.4+412255 & 16 35 11.45 +41 22 55.5 & 16 35 11.54 +41 22 57.4 & 0.72$\pm$0.09 & 0.94$\pm$0.02 \\ \hline
%% \end{tabular}
%% \end{center}
%% \end{table*}

%\begin{figure*}
%\includegraphics[width=0.95\textwidth]{figure14a}
%\caption{Postage stamps of all 15$\mu$m sources which have not been
%associated to any optical source. Also shown a 4\arcsec circle
%radius. (postage stamps from r'-band; 30\arcsec$\times$30\arcsec)}
%\end{figure*}

%\begin{figure*}
%\includegraphics[width=0.95\textwidth]{figure14b}
%\textbf{Figure 14}\textit{-- continued}
%\end{figure*}

\section{Summary}

The association of sources detected at 15$\mu$m in the \ELAIS N1 and
N2 areas with optical objects is presented. A 92\% of the sample
presents an optical identification to r'=24. The magnitude
distribution presents a maximum at r'=18 and a tail which extends to
fainter magnitudes. The distribution of point-like objects presents a
maximum one magnitude fainter.  The mid-IR to optical flux ratios,
$S_{15}/S_{r'}$, of the bright optical sources are in the range 1 to
$10^2$ and can be explained using simple models of cirrus, starbursts,
AGN and Arp220 spectral energy distributions. The tail of faint
objects, show larger $S_{15}/S_{r'}$, from $10^2$ to $10^3$ and can
only be explained assuming large luminosities and obscurations. Point
like objects show bluer g'-r' colour while higher mid-IR to optical
flux ratio than galaxies. The remaining 8\% of objects not identified
are faint in the mid-IR , with a $15\mu$m flux lower than 3\,mJy.
However, their mid-IR-to-optical flux ratio is larger than $10^3$
favouring the interpretation that they are associated with starbursts
or AGNs at high redshifts and highly obscured. The identification of
radio sources in the same areas is also presented. Their magnitude
distribution shows an increase on the number of sources towards faint
magnitudes. The number of unidentified objects is 44\%.

%The \ELAIS areas N1 and N2 will also be observed by the Spitzer Wide-area
%InfraRed Extragalactic Survey (SWIRE; Lonsdale et al. 2003). 

\section{Acknowledgements}

EAGS aknowledges support by EC Marie Curie Fellowship MCFI-2001-01809
and PPARC grant PPA/G/S/2000/00508.  The \ELAIS consortium also
acknowledges support from EC Training Mobility Research Networks 'ISO
Survey' (FMRX-CT96-0068) and 'Probing the Origin of the Extragalactic
background light (POE)' (HPRN-CT-2000-00138) and from PPARC.  This
paper is based on observations with \textit{ISO}, an ESA project, with
instruments funded by ESA Member States (especially the PI countries:
France, Germany, the Netherlands and the United Kingdom) and with
participation of ISAS and NASA. The INT is operated on the island of
La Palma by the Isaac Newton Group in the Spanish Observatorio del
Roque de los Muchachos of the Instituto de Astrofisica de Canarias.

%% \onecolumn
%% \begin{landscape}
%% \renewcommand{\arraystretch}{0.8}
%% \begin{center}
%% \begin{table*}
%% \caption{Optical properties of \ELAIS 15$\mu$m sources detected in N1 and N2.}
%% \label{tab:elais}
%% \begin{tabular}{LCCRRRRRRRRRRRL}
%% \hline\hline
%% \input{table15.tex}
%% \hline\hline
%% \end{tabular}

%% \end{table*}
%% \end{center}
%% \end{landscape}
%% \twocolumn

\end{document}